\documentclass{aastex63}
\usepackage[utf8]{inputenc}
\usepackage{placeins}
\usepackage{graphicx}
\usepackage{longtable}
\usepackage{amsmath}

\newcommand{\inv}{$^{-1}$ }
\newcommand{\invnospace}{$^{-1}$}

\begin{document}

\title{The Local Stellar Halo is Not Dominated by a Single Radial Merger Event}

\author{Thomas Donlon II}
\affiliation{Department of Physics, Applied Physics and Astronomy, Rensselaer Polytechnic Institute, 110 8th St, Troy, NY 12180, USA}
\correspondingauthor{Thomas Donlon II}
\email{donlot@rpi.edu}

\author{Heidi Jo Newberg}
\affiliation{Department of Physics, Applied Physics and Astronomy, Rensselaer Polytechnic Institute, 110 8th St, Troy, NY 12180, USA}

\author{Bokyoung Kim}
\affiliation{Department of Physics and Astronomy, Georgia State University, 25 Park Place, Suite 605, Atlanta, GA 30303, USA}
\affiliation{Institute for Astronomy, University of Edinburgh, Royal Observatory, Blackford Hill, Edinburgh EH9 3HJ, UK}

\author{Sebastien L\' epine}
\affiliation{Department of Physics and Astronomy, Georgia State University, 25 Park Place, Suite 605, Atlanta, GA 30303, USA}

\begin{abstract}
We use halo dwarf stars with photometrically determined metallicities that are located within 2 kpc of the Sun to identify local halo substructure. The kinematic properties of these stars do not indicate a single, dominant radial merger event (RME). The retrograde Virgo Radial Merger (VRM) component has [Fe/H] = -1.7. A second, non-rotating RME component we name Nereus is identified with [Fe/H] $=-2.1$ and has similar energy as the VRM. A possible third RME we name Cronus is identified that is co-rotating with the disk, has lower energy than the VRM, and has [Fe/H] = -1.2. We identify the Nyx Stream in the data. In addition to these substructures, we observe metal-poor halo stars ([Fe/H] $\sim$ -2.0 and $\sigma_v \sim 180$ km s\invnospace) and a Disk/Splash component with lower rotational velocity than the disk and lower metallicity than typically associated with the Splash. An additional excess of halo stars with low velocity and metallicity of [Fe/H] = -1.5 could be associated with the shell of a lower energy RME or indicate that lower energy halo stars have higher metallicity. Stars which comprise the “{\it Gaia} Sausage” velocity structure are a combination of the components identified in this work. 
\end{abstract}

\section*{} 

\twocolumngrid

\section{Introduction} \label{sec:intro}


Perhaps the most surprising discovery in the {\it Gaia} \citep{GaiaCollaboration2016} era is the proposal that the stellar halo is primarily composed of a single massive, ancient accretion event \citep{Deason2013,Belokurov2018}. This accretion event, discovered simultaneously by multiple groups as structures in velocity and chemical abundances, is known as the {\it Gaia} Sausage/\textit{Gaia}-Enceladus merger \citep[GSE,][]{Belokurov2018,Helmi2018}. 

GSE stars have a wide spread in radial velocity and near-zero rotational velocities, resulting in a ``sausage'' shape in plots of $v_r$ vs. $v_\phi$. This velocity profile is built up from symmetric $v_r$ velocity lobes caused by stars on high eccentricity orbits with an apogalacticon outside the solar circle \citep[e.g. Figure 3 of][]{Donlon2019}. Radial merger events (RMEs) produce shell structures at the apogalacticon of the orbits; inside of apogalacticon some stars are moving towards the Galactic center and some are moving away, so symmetric lobes in $v_r$ are a hallmark of a RME. These velocity lobes have been used to trace GSE debris on galaxy-wide scales \citep{Lancaster2019,Iorio2019,Necib2019,IorioBelokurov2021}. These cited works unanamously conclude that the halo is composed of a metal-poor, radially isotropic part (often called the ``in-situ'' halo), and a metal-rich, radially anisotropic part (the GSE).

The majority of previous literature is in agreement that the GSE debris fell into the proto-Milky Way 8-11 Gyr ago, and could be related to the co-formation of the Milky Way's (MW's) thick disk and inner stellar halo \citep{Helmi2018,Belokurov2018,Helmi2020}. 
However, \cite{Donlon2019} found that a young (2 Gyr ago) RME could explain a portion of the ``sausage'' velocity structure, and \cite{Donlon2020} determined that the distribution of RME debris shells in the stellar halo was consistent with a dwarf galaxy colliding with the Galactic center 2.7 Gyr ago. This recent RME is known as the Virgo Radial Merger (VRM).

\begin{figure*}
    \centering
    \includegraphics[width=\linewidth]{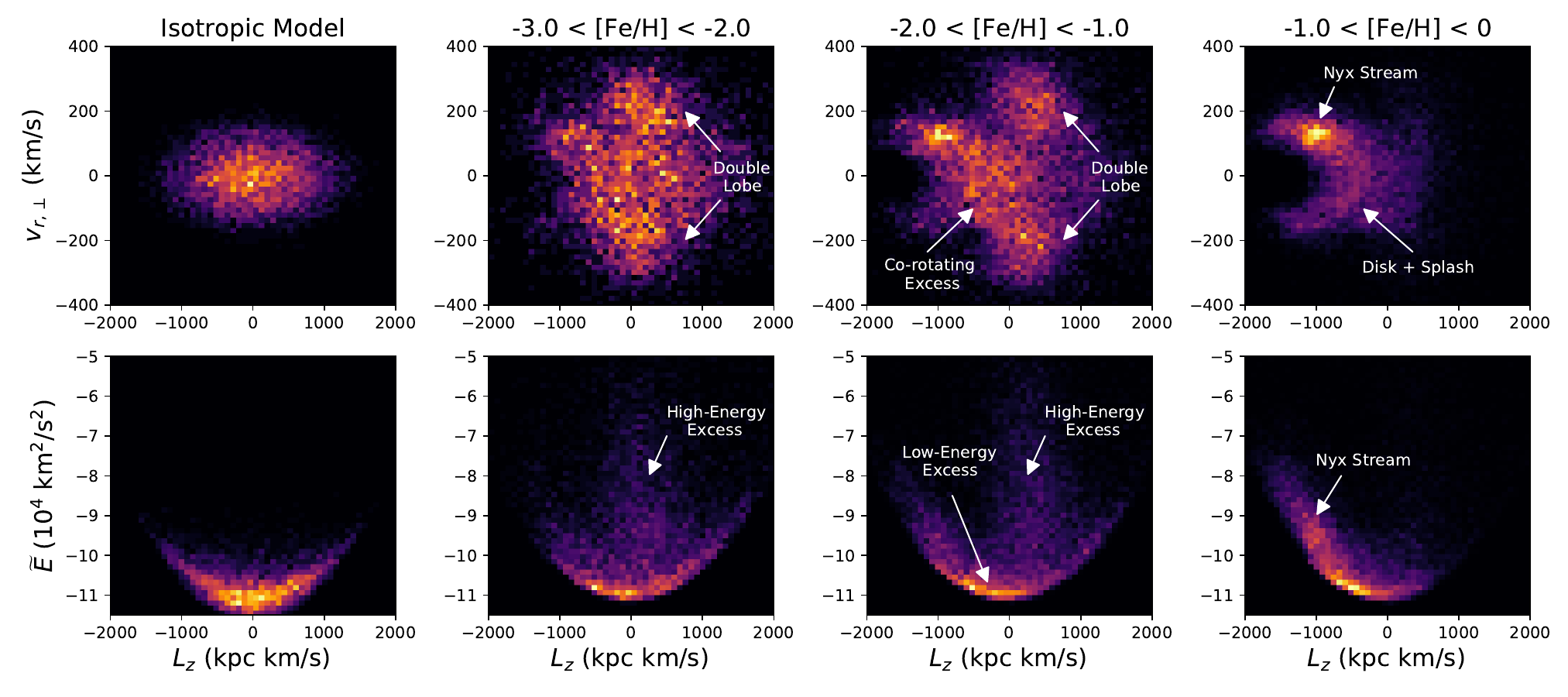}
    \caption{Heatmaps of kinematic properties of an isotropic halo model (left column) and the Gaia sample of local dwarf stars (right three columns, separated by metallicity as labeled at the top). The top row shows $v_{r,\perp}$ vs. $L_z$, and the bottom row shows $\widetilde{E}$ vs. $L_z$. Arrows and labels identify substructure. At low metallicities, the dwarf star population in the solar region appears similar to an isotropic halo background plus a non-rotating ($L_z\sim 0$) RME at higher energies. At intermediate metallicities the isotropic component is still evident, but at high energies the population is dominated by a retrograde RME, and at low energies there are an excess of prograde stars and an overdensity at $L_z$ = -1000 kpc km s\invnospace. At high metallicities, the dwarf stars are primarily made up of the $L_z$ = -1000 kpc km s\inv overdensity plus a crescent of disk/Splash stars bleeding up to positive $L_z$. The missing semicircle of data at $v_{r,\perp}=0$ and $L_z$ = -1500 kpc km s\inv is due to the proper motion cut that removes disk stars.}
    \label{fig:kinematics}
\end{figure*}

From local halo stars, we show that VRM stars ([Fe/H] = -1.7) have a net retrograde motion. Stars with low $|v_r|$ have net prograde motion, indicating a distinct substructure. Additionally, there is a non-rotating, metal-poor ([Fe/H] = $-2.1$) radial component that is consistent with either a single large RME or many minor RMEs. These components\footnote{\cite{Malhan2022} was published while this work was publically available and undergoing the review process. If one assumes that their study of objects in the entire halo is directly comparable to the distribution of stars near the Sun, the LMS-1/Wukong and Pontus structures identified by \cite{Malhan2022} are possibly related to the Nereus and Cronus structures we identify in this work.} all contribute to the {\it Gaia} Sausage velocity structure. 

\section{Dwarf Stars in the Local Solar Neighborhood} \label{sec:catalog}

\subsection{The Catalog}

We select 296,933 dwarf stars in the local stellar halo, with proper motions and parallaxes from the \textit{Gaia}-EDR3 survey \citep{GaiaEDR3}, using the criterion given in $\S$2.3.1 of \cite{Kim2020} and a parallax zero-point of -0.017 mas. 71\% (99\%) of these stars are located within 1 (2) kpc of the Sun. Photometric metallicity estimations for each star are obtained from \cite{KimLepine2021}.

\subsection{Coordinate System \& Notation}

The majority of the stars in our sample are missing line-of-sight velocity measurements that are required to calculate full 3D velocities. However, if one selects only stars that are located within 20$^\circ$ of the Galactic poles, then the tangential velocities of each star correspond to $v_{x,\perp}$ and $v_{y,\perp}$ \citep[e.g.][]{Morrison1990,Kordopatis2017,Kim2020,Fernandez-Alvar2021}. This ``polar'' sample contains 36,410 dwarf stars. The ``$\perp$'' subscript on a velocity indicates that it was calculated without line-of-sight velocity information. 

In the solar neighborhood, radial velocity $v_r$ and rotational velocity $v_\phi$ are roughly equivalent to $-v_x$ and $v_y$, respectively, so this work uses these quantities interchangeably. We use a right-handed coordinate system: positive $v_x$ is in the direction of the Galactic center, and positive $v_y$ is in the direction of the disk's rotation, so disk stars have negative $L_z$. We use the Sun's motion from \cite{Hogg2005} to transform from heliocentric velocities to Galactocentric velocities.

\begin{figure*}
    \centering
    \includegraphics[width=\linewidth]{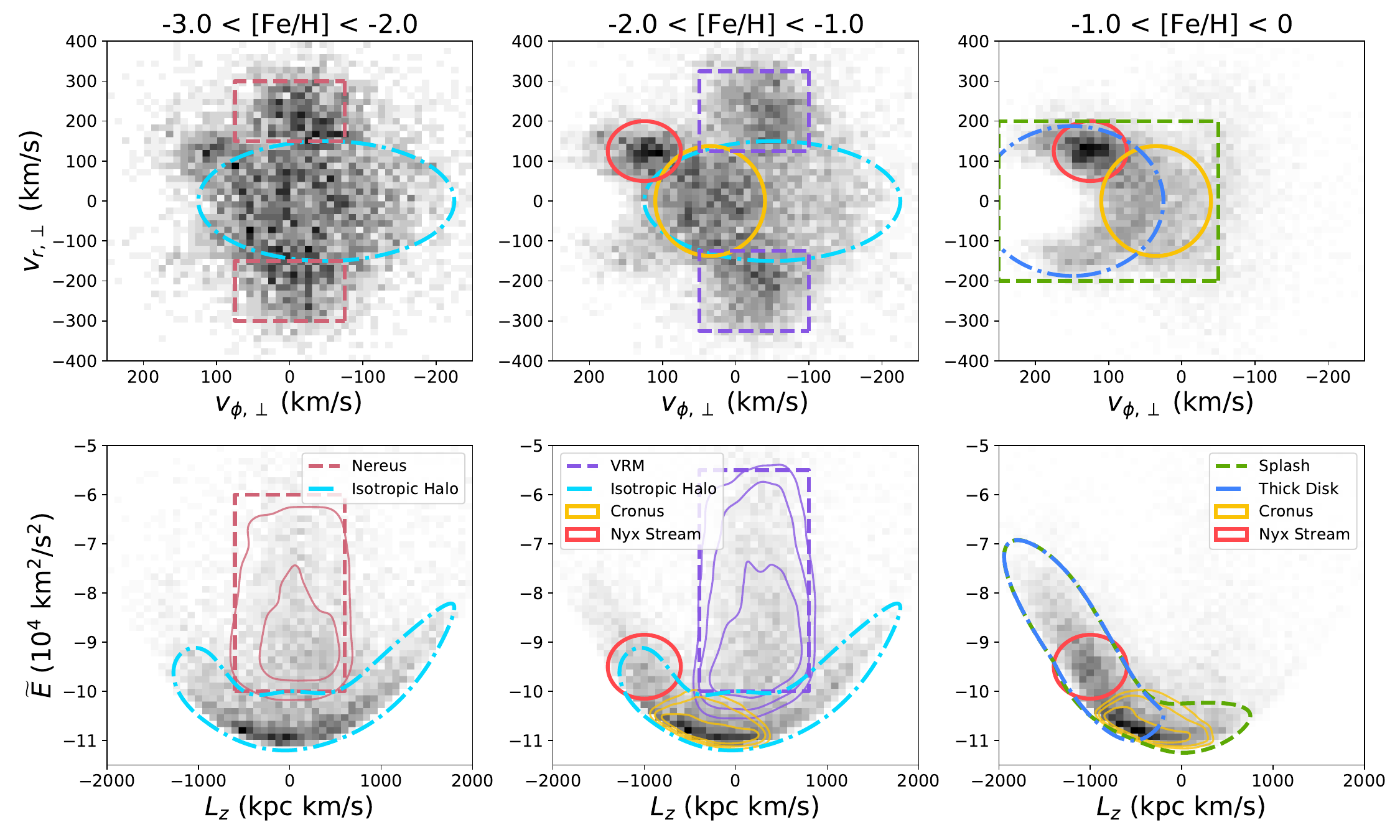}
    \caption{Kinematic properties of the various structures identified in this work. The dark red and purple dashed lines indicate the extent of the non-rotating and retrograde high-energy RMEs, respectively. The solid yellow line indicates the extent of the low energy RME, and the dashed light-blue line surrounds the isotropic halo background. The light red ellipse indicates the location of the Nyx Stream, the green and blue dashed lines show the extent of Splash and disk stars, respectively. Contours in the bottom row show the 95\%, 75\%, and 50\% levels of stars within the VRM, Nereus, and Cronus selections in the top row. The stars in the center of the top panels have low velocity, and therefore lower energy since the stars in the solar neighborhood have similar potential energies.}
    \label{fig:regions}
\end{figure*}

We calculate pseudo-energy, $\widetilde{E}$, as: \begin{equation}
\widetilde{E} = \Phi(\vec{r}) + \frac{1}{2}\left(v^2_{x,\perp} + v^2_{y,\perp}\right),
\end{equation} where $\Phi(\vec{r})$ is the potential model used in \cite{Donlon2019}. Assuming that the missing velocity component is small, $\widetilde{E} \approx E$. Similarly, our calculations of $L_z$ are missing line-of-sight velocity contributions, but we choose not to alter this notation since ``$L_\perp$'' already represents a particular kinematic quantity. 

\subsection{Isotropic Halo Background}

In the absence of halo substructure, one might imagine the distribution of halo star kinematics would resemble an isotropic Plummer profile \citep{Plummer1911}. We randomly sampled a Plummer profile with a scale radius of 12 kpc and mass of 10$^{11}$ M$_\odot$ \citep[reasonable estimates for the inner MW halo;][]{Weiss2018b,PostiHelmi2019} centered on the Galactic center, gave each particle a randomly oriented velocity, and then removed particles outside the 3D footprint of the dwarf data, following the procedure outlined in \cite{SheltonInPrep}. The resulting model (Figure \ref{fig:kinematics}) provides an estimate for the kinematics of local halo dwarfs under the assumption that the stellar halo was built up by the accretion of many phase-mixed progenitors on random initial orbits. Any energy offset between the isotropic model and the dwarf data can be resolved by tuning the scale length and mass of the isotropic model.

\vspace{2cm}

\subsection{Kinematics of Nearby Dwarf stars}

The {\it Gaia} Sausage is easily identifiable in $v_r$ and $v_\phi$ ($L_z$) in the local stellar halo. Therefore, we look for halo structure using these quantities. Figure \ref{fig:kinematics} shows the distributions of $v_{r,\perp}$, $\widetilde{E}$, and $L_z$ in three metallicity ranges in the polar dwarf star sample, compared to the isotropic halo model.

Note the high energy ``plumes'' at $L_z \sim 0$ kpc km s\inv in the lowest metallicity bin, and at $L_z \sim 400$ kpc km s\inv in the intermediate metallicity bin. These plumes correspond to the $v_{r,\perp} \sim \pm200$ km s\inv lobes in the lowest metallicity bin and the slightly retrograde lobes of stars with $v_{r,\perp} \sim \pm220$ km s\inv in the intermediate metallicity dwarf stars, respectively. In addition to these two sets of lobes, there is an excess of stars with negative $L_z$ at low $|v_{r,\perp}|$ at intermediate metallicity. These stars are all typically attributed to the {\it Gaia} Sausage, but the variation in angular momentum, energy, and metallicity leads one to believe that the characteristic ``sausage'' shape is built up from multiple substructures.

At primarily high metallicities there is an excess of stars at $(L_z, v_{r,\perp}) = $(-1000 kpc km s\invnospace, 150 km s\invnospace) that is not mirrored at negative $v_{r,\perp}$. This feature is consistent with the Nyx Stream \citep{Necib2020}, which we discuss in Section \ref{sec:stream}.


The low and intermediate metallicity dwarf stars contain stars that are distributed similarly to an isotropic background stellar halo (light blue in Figure \ref{fig:regions}), creating tension with the \cite{Naidu2020} claim that the stellar halo is entirely composed of substructure. It is consistent, however, with reports that the metal-poor stellar halo is fit well by a radially isotropic model  (\citealp{Lancaster2019,Iorio2019,Necib2019,IorioBelokurov2021}, but see \citealp{Cunningham2019}). The analysis of \cite{Naidu2020} focused on stars beyond the solar circle, so our findings may indicate that the inner halo is smoother than the outer halo.


At values of $L_z > 1000$ kpc km s\invnospace, the low metallicity dwarf stars extend upwards to higher energies and higher $L_z$ than the model. This retrograde component is possibly related to the Sequoia structure \citep{Myeong2019}. At values of $L_z < -1000$ kpc km s\inv and high metallicity, the stars extend to higher energy than the isotropic model. This could result from uncertainties in the dwarf star metallicity estimates, thick disk and Splash \citep{Belokurov2020} stars extending to metallicities lower than [Fe/H] = -1.0, some dynamical process between the inner stellar halo and the disk, or the actual stellar halo not being isotropic.

\section{Multiple Radial Merger Events} \label{sec:rmes}

In this section and Section \ref{sec:stream}, we analyze the components of the stellar halo, as identified in Figure \ref{fig:regions}. 

\subsection{High-Energy Structures}

\begin{figure}
    \centering
    \includegraphics[width=\linewidth]{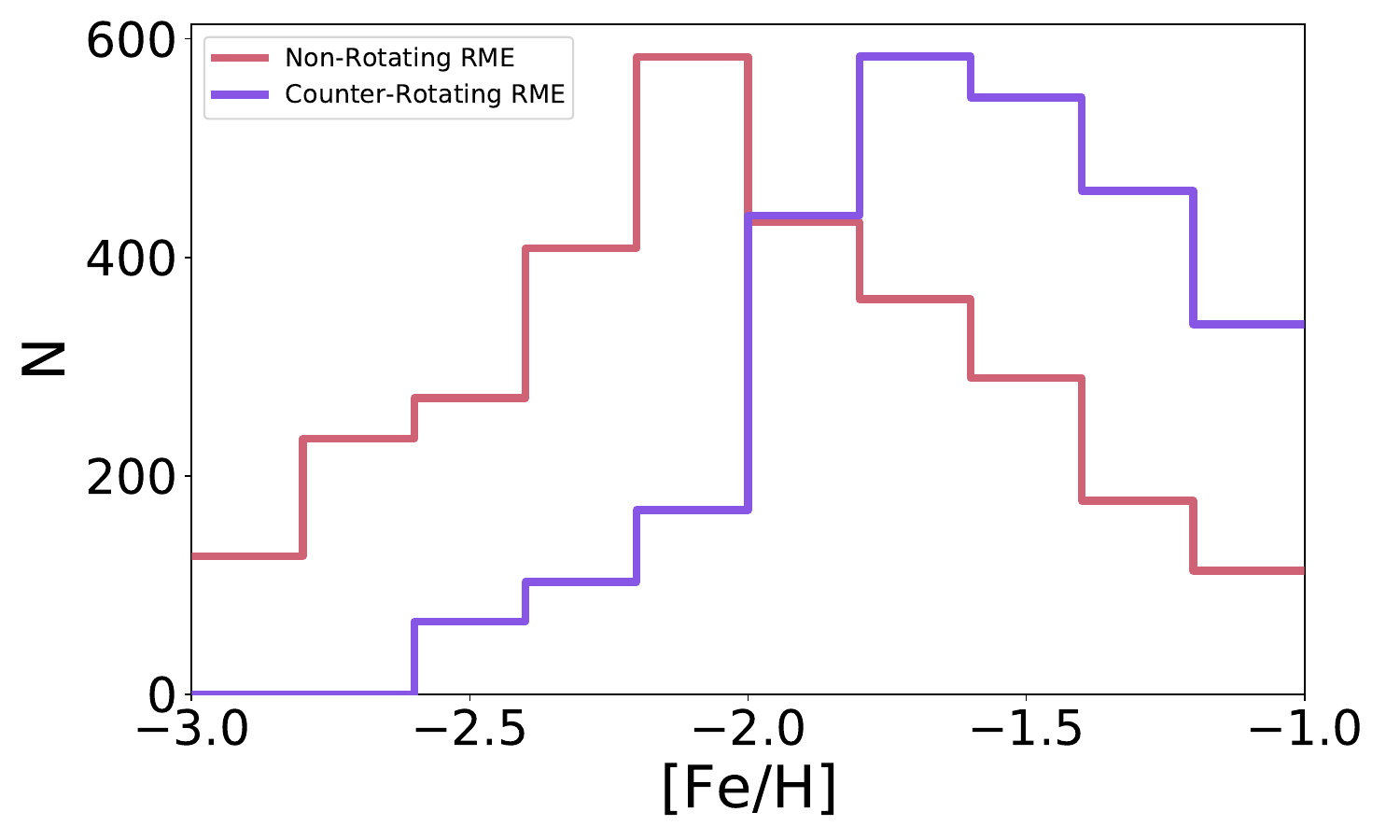}
    \caption{Fit metallicity distribution functions (MDFs) for the two high-energy RMEs. The purple line corresponds to the VRM, and the red line corresponds to a low metallicity, non-rotating (collection of) RME(s). The two structures contribute roughly similar numbers of stars to the nearby stellar halo.}
    \label{fig:rme_mdfs}
\end{figure}

It is unexpected that the high energy plumes of stars have different $L_z$ in the low and intermediate metallicity samples. If the high-energy stars all belonged to the same progenitor, then they would have been moving together immediately before accretion, and should have similar angular momenta regardless of their metallicity. This shift does not appear to be explained by an internal rotation in the progenitor dwarf galaxy: \cite{Naidu2021} model a merger between a disky GSE progenitor and a proto-MW, and predict that GSE stars with $L_z\sim0$ kpc km s\inv should have a higher metallicity than GSE stars with nonzero $L_z$, which is the opposite of our observations. 

\begin{figure*}[!p]
    \centering
    \includegraphics[width=0.95\linewidth]{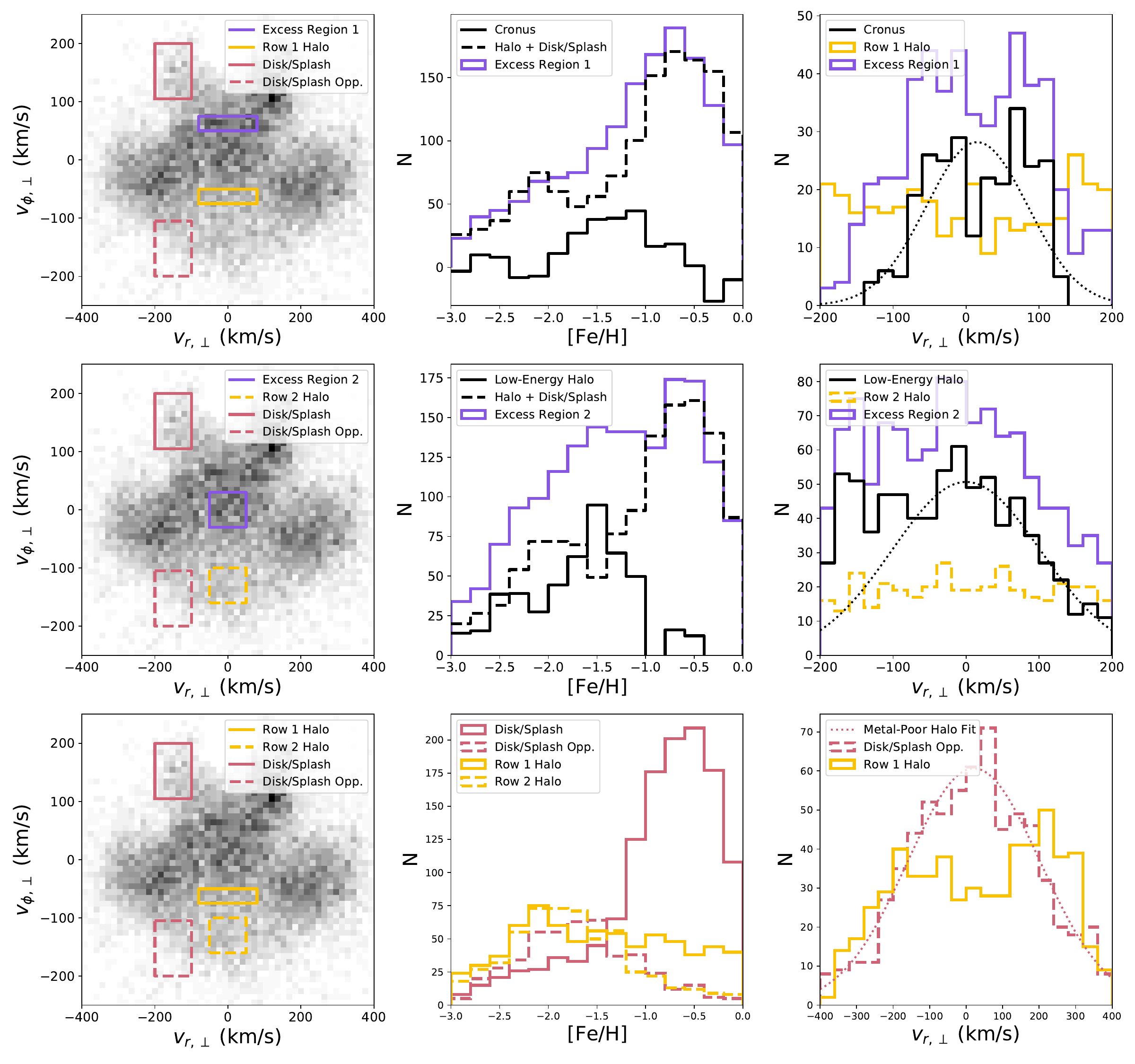}
    \caption{Metallicity and radial velocity of the low-energy structures. The top, middle, and bottom rows present results for the low energy RME (aka Cronus), the low energy halo, and the metal-poor halo, respectively. Purple boxes show two samples in regions where an excess has been identified, yellow boxes the halo, and red boxes the disk (solid line) and a region with opposite velocity (dashed line). The middle panels in the top and middle rows show the MDFs of the overdense region (purple line), the estimated total background (dashed black line), and the difference between them (solid black line, corresponding to Cronus and the Low Energy Halo). The bottom middle panel shows the MDF of the halo and disk comparison regions. The rightmost column shows the $v_{r,\perp}$ distributions for each substructure, using only stars with -1.5 $<$ [Fe/H] $<$ -1. The top right panel shows the radial velocity distributions for stars with the $v_\phi$ range of the Cronus excess region (purple line) and the halo region with an opposite $v_{\phi,\perp}$ (yellow line). The yellow distribution is artificially high at large $|v_{r,\perp}|$ due to the VRM velocity lobes. The solid black line shows the difference between the purple and yellow distributions, which represents the Cronus $v_{r,\perp}$ distribution. A dashed black line shows a Gaussian that has been normalized to the solid black distribution; the depletion near $v_{r,\perp}=0$ in the solid black distribution compared to the Gaussian suggests that Cronus is associated with a RME. The right panel in the center row shows the velocity distribution of the low energy halo; the Gaussian is not fit on the left side due to apparent pollution of the data by VRM stars. The bottom right panel shows the radial velocity distribution of halo stars within the $v_\phi$ range of the dashed red box, and a Gaussian fit to it. Also shown is the $v_{r,\perp}$ distribution for stars with the $v_\phi$ range of the solid yellow box, which includes halo in the center plus VRM lobes on each side.}
    \label{fig:lowe_rme}
\end{figure*}

Motivated by this discrepency, we selected stars from the polar sample with $\widetilde{E}>-9\times 10^4$ km$^2$ s$^{-2}$ and $|L_z|<1000$ kpc km s\invnospace, binned them in 0.2 dex wide metallicity bins, and then optimized a two-component Gaussian mixture model $f(L_z)$ on the $L_z$ distributions of the data in each bin. The amplitude of each Gaussian was fit independently for each $L_z$ bin, but the position and dispersion of each Gaussian were simultaneously fit to all bins. These mixture models were optimized by minimizing a residual sum of squares ($RSS$) using \verb!scipy! \citep{scikit-learn}, given by \begin{equation}
RSS = \sum_i^N \left(\int_{L_{z,i}}^{L_{z,i+1}} f(L_z)\;\textrm{d}L_z - n_i\right)^2,
\end{equation} where $N$ is the number of bins in the histogram, $L_{z,i}$ and $L_{z,i+1}$ give the bounds of the $i$th bin, and $n_i$ is the number of stars in the $i$th bin of the data histogram.

The optimization favors a [Fe/H]$=-2.1$ component with $L_z$ = 50 kpc km s\invnospace, and a retrograde [Fe/H]$=-1.7$ component with $L_z$ = 380 kpc km s\invnospace. The MDFs of these components are shown in Figure~\ref{fig:rme_mdfs}. The retrograde component's MDF appears similar to that of a luminous dwarf galaxy \citep{Kirby2013}, and matches the metallicity of the Virgo Overdensity \citep{Duffau2006,An2009}. Therefore we identify the $L_z\sim380$ kpc km s\inv component as the VRM. We name the lower metallicity, low-rotation component Nereus.

While one assumes the counter-rotating VRM results from a single merger event, the low $L_z$ radial merger component is arguably non-rotating, and could result from one progenitor or a group of smaller mergers with a range of $L_z$. The similar energy of the stars in Nereus and the VRM suggests that they were accreted at roughly the same time as one another; Nereus could possibly have been related to the VRM before they were accreted. These RMEs contribute similar numbers of stars to the local stellar halo, which may hold clues regarding their formation.

\subsection{Low-Energy Sample}

\begin{figure*}
    \centering
    \includegraphics[width=0.9\linewidth]{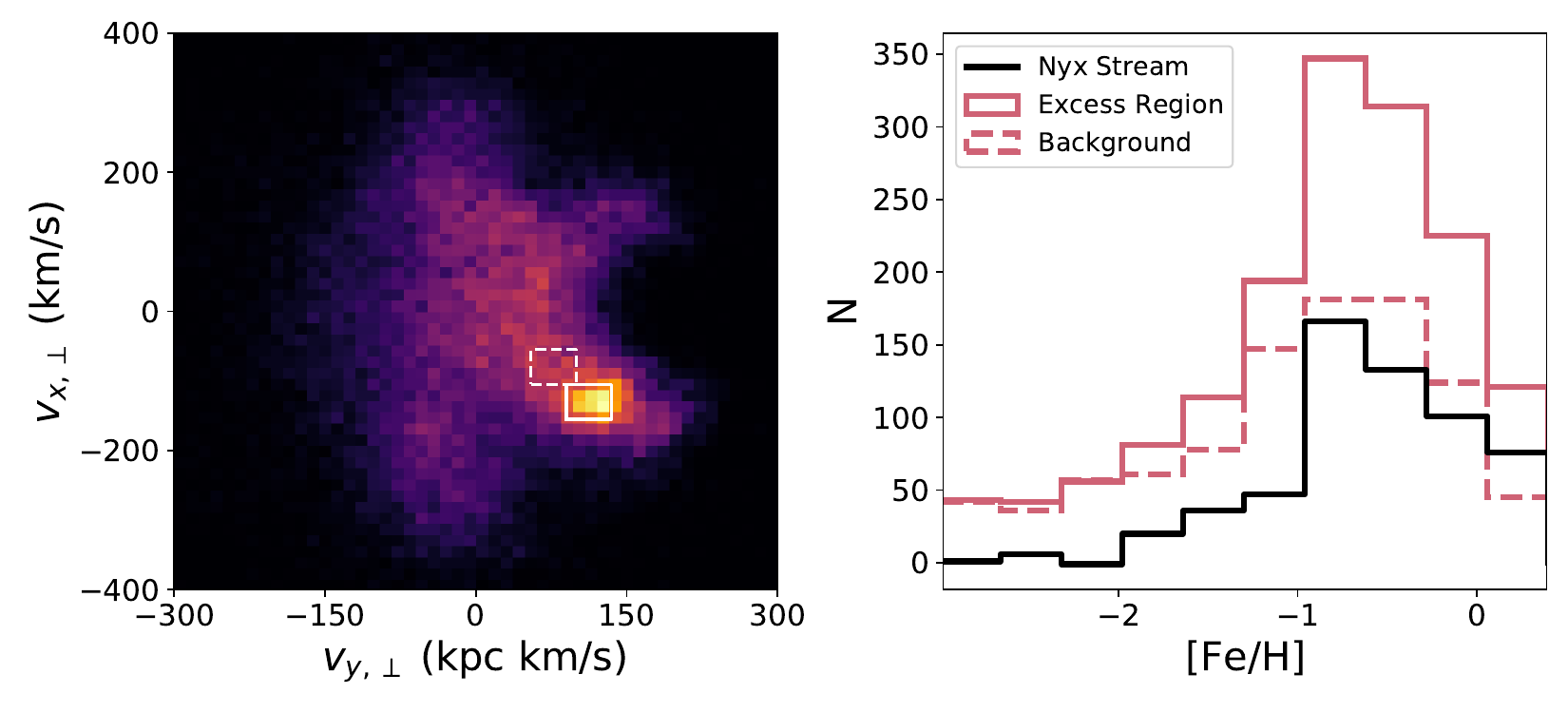}
    \caption{Properties of the Nyx Stream. The left panel shows a heatmap of velocities in our polar sample. A solid white outline marks the velocity extent of the excess that has been associated with the Nyx Stream. A dashed white outline marks an adjacent background region of similar size. The right panel shows the MDF for the excess region, the background region, and their difference which represents the distribution of the Nyx Stream stars.}
    \label{fig:stream}
\end{figure*}

The low and intermediate metallicity polar dwarf stars contain an excess of co-rotating stars centered on $(L_z, v_{r,\perp})=(-400,0)$. We cannot fit a mixture model as we did to the high energy data because the distribution in $L_z$ is affected by energy constraints imposed by the gravitational potential in the Solar neighborhood. 

Instead, we identify excesses in the central (lower energy) regions of the $(v_{\phi, \perp}, v_{r,\perp})$ heatmap, as identified by the purple boxes in Figure~\ref{fig:lowe_rme}. We plot the metallicity distribution of the stars in that velocity region, and then subtract off the expected distributions from the halo (yellow boxes) and disk (red box minus the red dashed box). The number of disk stars in the purple region is calculated from the ratio of stars in the purple-minus-yellow to red-minus-red-dashed boxes at high metallicity (-1$<$[Fe/H]$<$0), multiplied by the number of disk stars in the red-minus-red-dashed boxes in a particular metallicity bin. The number of halo stars in the purple region is calculated from the ratio of stars in the purple region to stars in the yellow region at low metallicity (-3$<$[Fe/H]$<$-2), multiplied by the number of stars in the yellow box in a each metallicity bin.

Figure~\ref{fig:lowe_rme} shows that the metal-poor halo stars with $-200 < v_{\phi, \perp} < -100$ km s\inv have a peak metallicity of -2.0 and a velocity dispersion of 180 km s\invnospace. Excesses of co-rotating and non-rotating stars at low energy have intermediate metallicities of $\sim -1.2$ and $\sim -1.5$, respectively. The $v_{r, \perp}$ distribution of the co-rotating stars shows the double Gaussian profile associated with a RME; the distribution is inconsistent with a single Gaussian at the p$<$0.01 level as per Hartigan's Dip Test \citep{diptest}. The small shift of the double-lobed distribution towards positive $v_{r,\perp}$ is a consequence of the fact that shells are constantly moving outward from the Galactic center \citep{Sanderson2013}. The stars in this low energy, double-lobed distribution are likely associated with yet a third RME remnant, which we name Cronus.

It is unclear what causes the Low Energy Halo feature at zero velocity. It is possible that the overall halo distribution is simply higher metallicity at lower energy, or the Low Energy Halo could be a separate halo component formed from one or more merger events. 

\section{Nyx Stream} \label{sec:stream}

Using deep learning techniques, \cite{Necib2020} found a collection of co-moving stars near the Sun in {\it Gaia} data. They claimed that this was an accreted stellar stream, which they named the Nyx Stream. This structure was corroborated by \cite{ReFiorentin2021}, who called it the Icarus Stream, and claimed it could be associated with the GSE. Recent findings suggest that Nyx was not accreted, and is related to the MW's thick disk \citep{Zucker2021}.

The polar dwarf star data in Figure \ref{fig:kinematics} contains a prominent overdensity around $(L_z, v_{r,\perp}) = $ (-1000 kpc km s\invnospace, 150 km s\invnospace) in intermediate and high metallicity stars. 
Figure \ref{fig:stream} shows the MDF for the structure, which peaks at [Fe/H] $= -0.8$ (\citealp{Necib2020} found [Fe/H] = $-0.5$). A Kolmogorov-Smirnov test \citep{kstest} indicates the MDF of the Nyx excess region is inconsistent with that of the background with a p-value of $3.2 \times 10^{-6}$, supporting the idea that the stars in this co-moving group are different from those with adjacent velocities.

The Nyx Stream has the same energy as Group D from \cite{Donlon2019}, which is potentially related to the HAC Group from \cite{Gryncewicz2021} and is thought to be part of a RME. It is possible that this stellar stream is related to those structures.

\section{Conclusions} \label{sec:conclusions}

Using the kinematics and metallicities of {\it Gaia} EDR3 dwarf stars within 2 kpc of the Sun, we have determined that the local stellar halo is composed of multiple halo components, none of which dominate the sample. These components include:

\begin{itemize}
\item \textit{VRM:} A retrograde RME with peak metallicity [Fe/H]$= -1.7$ and $L_z\sim380$ kpc km s\invnospace. This structure has a double-lobed $v_r$ velocity profile (typical for a RME) with $|v_{r, \perp}|=220$ km s\invnospace.

\item \textit{Non-rotating RME (Nereus):} A collection of stars with energies similar to the VRM, but with lower metallicity of [Fe/H] = $-2.1$, $|v_{r, \perp}|=200$ km s\invnospace, and near zero rotational velocity. We argue from its energy that this formed from a single, recent RME, but it could be made up of one or more satellites that fell into the MW at early times and were radialized over time. We name this component Nereus. 

\item \textit{Low-Energy RME (Cronus):} This collection of prograde ($L_z\sim-400$ kpc km s\invnospace), low-energy stars with [Fe/H] = -1.2 also has a double-lobed velocity profile that is characteristic of a RME, with $v_{r, \perp}=-40,+90$ km s\invnospace. We name this halo component Cronus.

\item \textit{Nyx Stream:} An excess of stars with $(v_{x,\perp}, v_{y,\perp}=(-130, 112.5)$ km s\invnospace and a metallicity of [Fe/H] = -0.8.

\item \textit{Metal-Poor Halo:} A distribution of stars with metallicity [Fe/H] $\sim -2.0$, and a velocity dispersion of 180 km s\invnospace.

\item \textit{Low Energy Halo:} An excess of low energy stars with [Fe/H]$\sim -1.5$, near zero velocity, and a velocity dispersion of 100 km s\invnospace. It is unclear whether these stars are associated with the shell of a RME or whether they indicate that the metallicity of the halo increases at low energy.

\item \textit{Disk/Splash:} A group of primarily co-rotating stars that have disk metallicity but appear to extend to metallicities of [Fe/H]$<-1.0$, which is lower than is typically attributed to the Splash. These stars also have higher velocities than is typical for the disk. 

\end{itemize}

The GSE structure is typically claimed to be the dominant component of the stellar halo. Our results indicate that the GSE stars are composed of multiple substructures. Kinematic cuts for the GSE typically select all low-$v_\phi$ and/or all high-eccentricity objects. These selections will contain stars from the VRM, Nereus, Cronus, the Metal-Poor Halo, and the Low Energy Halo. The multiple structures present in this kinematic cut might explain the discrepancies between dynamical and chemical cuts used to select the GSE \citep{Buder2021}.

\acknowledgments
This work was supported by NSF grant AST 19-08653; contributions made by Manit Limlamai; and the NASA/NY Space Grant.

This work has made use of data from the European Space Agency (ESA) mission
{\it Gaia} (\url{https://www.cosmos.esa.int/gaia}), processed by the {\it Gaia}
Data Processing and Analysis Consortium (DPAC,
\url{https://www.cosmos.esa.int/web/gaia/dpac/consortium}). Funding for the DPAC
has been provided by national institutions, in particular the institutions
participating in the {\it Gaia} Multilateral Agreement.

\bibliographystyle{aasjournal}
\bibliography{references.bib}

%
%
%

\end{document}